\documentclass[10pt,conference]{IEEEtran}
\IEEEoverridecommandlockouts
\usepackage{cite}
\usepackage{amsmath,amssymb,amsfonts,url}
\usepackage{graphicx}
\usepackage{textcomp}
\usepackage{xcolor}
\usepackage{comment}
\usepackage[nonumberlist]{glossaries}
\usepackage{multirow}
\usepackage[nonumberlist]{glossaries}
\usepackage{threeparttable}
\usepackage{graphicx} 
\usepackage{tikz}
\usepackage{pgfplots}
\pgfplotsset{compat=1.18}
\usepackage{subcaption}
\usepackage{acronym}
\usepackage{booktabs}
\usepackage{algorithm}
\usepackage{algpseudocode}
\usepackage[algo2e]{algorithm2e} 

\usepackage{xcolor} 
 \usepackage{float}
\def\BibTeX{{\rm B\kern-.05em{\sc i\kern-.025em b}\kern-.08em
    T\kern-.1667em\lower.7ex\hbox{E}\kern-.125emX}}

\newacronym{2D}{2D}{Two Dimensional}
\newacronym{6G}{6G}{Sixth Generation}
\newacronym{AI}{AI}{Artificial Intelligence}
\newacronym{BCE}{BCE}{Binary Cross-Entropy}
\newacronym{BER}{BER}{Bit Error Rate}
\newacronym{BLER}{BLER}{Block Error Rate}
\newacronym{CDL}{CDL}{Clustered Delay Line}
\newacronym{CNN}{CNN}{Convolutional Neural Network}
\newacronym{CPU}{CPU}{Central Processing Unit}
\newacronym{CSI}{CSI}{Channel State Information}
\newacronym{DL}{DL}{Deep Learning}
\newacronym{DMRS}{DMRS}{Demodulation Reference Signal}
\newacronym{FFT}{FFT}{Fast Fourier Transform}
\newacronym{FLOP}{FLOP}{Floating-Point Operation}
\newacronym{GFLOP}{GFLOP}{Giga Floating-Point Operation}
\newacronym{FLOPs}{FLOPs}{Floating-Point Operations}
\newacronym{GPU}{GPU}{Graphics Processing Unit}
\newacronym{GNN}{GNN}{Graph Neural Network}
\newacronym{IFFT}{IFFT}{Inverse Fast Fourier Transform}
\newacronym{LN}{LN}{Layer Normalization}
\newacronym{LDPC}{LDPC}{Low-Density Parity-Check}
\newacronym{LLR}{LLR}{Log-Likelihood Ratio}
\newacronym{LLM}{LLM}{Large Language Model}
\newacronym{LMMSE}{LMMSE}{Linear Minimum Mean Squared Error}
\newacronym{LOS}{LOS}{Line-of-Sight}
\newacronym{LS}{LS}{Least-Squares}
\newacronym{ML}{ML}{Machine Learning}
\newacronym{MHSA}{MHSA}{Multi-Head Self-Attention }
\newacronym{MIMO}{MIMO}{Multiple-Input-Multiple-Output}
\newacronym{NLOS}{NLoS}{Non-LoS}
\newacronym{NPU}{NPU}{Neural Processing Unit}
\newacronym{OFDM}{OFDM}{Orthogonal Frequency Division Multiplexing}
\newacronym{OOM}{OOM}{out-of-memory}
\newacronym{PHY}{PHY}{Physical Layer}
\newacronym{PTQ}{PTQ}{Post-Training Quantization}
\newacronym{QAM}{QAM}{Quadrature Amplitude Modulation}
\newacronym{QAT}{QAT}{Quantization-Aware Training}
\newacronym{RAN}{RAN}{Radio Access Networks}
\newacronym{RB}{RB}{Resource Block}
\newacronym{RG}{RG}{Resource Grid}
\newacronym{SIMO}{SIMO}{Single-Input-Multiple-Output}
\newacronym{SISO}{SISO}{Single-Input-Single-Output}
\newacronym{SNR}{SNR}{Signal-to-Noise-Ratio}
\newacronym{STE}{STE}{Straight‑Through Estimator}
\newacronym{UE}{UE}{User Equipment}
\newacronym{URLLC}{URLLC}{Ultra-Reliable Low-Latency}
\newacronym{V2X}{V2X}{Vehicle-to-Everything}
\newacronym{ZF}{ZF}{Zero-Forcing}

\begin{document}

\title{{Computationally Efficient Neural Receiver via Axial Self-Attention}}

\author{
    \IEEEauthorblockN{
        SaiKrishna Saketh Yellapragada\IEEEauthorrefmark{1}\thanks{Corresponding author: saikrishna.yellapragada@aalto.fi. 
        The work of the first author has been supported in parts by the Research Council of Finland (grant no. 359848) and the European Union's 6GARROW project (No. 101192194).},
        Atchutaram K. Kocharlakota \IEEEauthorrefmark{3}\thanks{\IEEEauthorrefmark{3}Work done while the author was affiliated with Aalto University.},
        Mário Costa\IEEEauthorrefmark{2},
        Esa Ollila\IEEEauthorrefmark{1},
        Sergiy A. Vorobyov\IEEEauthorrefmark{1}
    }
    \IEEEauthorblockA{
        \IEEEauthorrefmark{1}\textit{Aalto University, Finland} \hspace{1em}
        \IEEEauthorrefmark{2}\textit{Nokia, Portugal}
    }
}

\maketitle
\begin{abstract}
Deep learning-based neural receivers offer promising physical-layer performance for next-generation wireless systems. We propose an axial self-attention transformer-based neural receiver that achieves superior \gls{BLER} performance compared with state-of-the-art receivers, while significantly improving computational efficiency during both inference and large-scale training. By factorizing attention operations along temporal and spectral axes, the proposed architecture reduces computational complexity from $O((TF)^2)$ to $O(T^2F+TF^2)$, yielding substantially fewer floating-point operations and attention matrix multiplications per transformer block. Experimental validation under 3GPP \gls{CDL} channels demonstrates consistent performance gains across varying mobility scenarios. Under non-line-of-sight conditions, our proposed axial neural receiver outperforms global self-attention and convolutional neural receiver baselines at 10\% BLER and 1\% BLER respectively, with reduced computational complexity. 
\end{abstract}

\section{Introduction}
\label{sec:intro}
As wireless communications advance toward \gls{6G} \gls{RAN}, \gls{DL}-based neural receivers are emerging as promising \gls{PHY} solutions that can jointly learn channel estimation, equalization, and soft demapping directly from received \gls{OFDM} \glspl{RG}. 3GPP Release~20 positions \gls{AI} as an important enabler for future air-interface and network-intelligence evolution. However, deploying neural receivers in real-time systems remains challenging due to stringent latency and compute budgets, especially for large time-frequency \glspl{RG}.

\gls{CNN}-based neural receivers jointly optimize channel estimation, equalization, and demapping by training a single architecture to map received signals directly to  \glspl{LLR}~\cite{deeprx,e2e_faa,trainable_seb}. Extensions to \gls{MIMO} have been proposed in~\cite{nrx_seb,deeprx_mimo_icc}, leveraging convolutional layers to capture time--frequency correlations as well as \gls{GNN}-based modules to mitigate multi-user interference. Recent studies have shown that \gls{CNN}-based neural receivers exhibit notable resilience to ultra-low bit quantization when subjected to model efficiency techniques such as \gls{QAT} and \gls{PTQ} \cite{saketh_qat,saketh_asilomar25}. Consequently, neural receivers represent a promising solution for deployment at the hardware-constrained 6G network edge.

Transformer architectures have achieved remarkable success in various domains including natural language processing and computer vision, especially with \glspl{LLM}, motivating their exploration for wireless communication \cite{vaswani2017attention,unified_transformer,atchutJournal,tingtingGAN}. In transformers, the \gls{MHSA} mechanism enables global context modeling by computing attention across all positions in the input sequence, providing key benefits for wireless communications where channel responses exhibit dependencies across both time and frequency domains due to multipath propagation and Doppler effects. The authors of \cite{unified_transformer} demonstrated effective \gls{OFDM} \gls{RG} processing by applying \gls{MHSA} to non-overlapping \glspl{RB} with \gls{2D} positional encodings that capture time-frequency dependencies. 
When applied to \gls{2D} time--frequency grids, standard \gls{MHSA} flattens the resource grid into a single sequence, yielding a computational complexity of \(\mathcal{O}((TF)^2)\), where \(T\) and \(F\) denote the temporal and spectral dimensions of the processed grid, respectively~\cite{unified_transformer}. Although~\cite{unified_transformer} mitigates this cost by operating on small non-overlapping tiles with \(T=14\) symbols and \(F=12\) subcarriers, practical systems typically require much larger frequency dimensions, with \(F\) often one or two orders of magnitude larger. Consequently, the quadratic scaling of standard \gls{MHSA} becomes a computational bottleneck for modern \gls{OFDM} systems with large time--frequency bandwidth parts.

To address these limitations, we draw inspiration from axial attention in computer vision~\cite{ho2019axial,axialpanoptic}, whose factorized design aligns naturally with the separable time-frequency correlation structure of wireless channels \cite{molisch2011wireless}. Building on this insight, we propose an axial-attention neural receiver that applies self-attention sequentially along the time and frequency axes.This reduces the computational complexity of the transformer neural receiver to $\mathcal{O}(T^2F + TF^2)$ while preserving the ability to capture long-range temporal and spectral dependencies across large \glspl{RG}. By mitigating the quadratic cost of standard \gls{MHSA}, the proposed axial neural receiver enables energy-efficient, low-latency inference suitable for \gls{AI}-\gls{RAN} in \gls{6G}. Moreover, by factorizing attention along the time and frequency axes, the proposed architecture reduces the computational burden of both training and inference, making it more practical for development and deployment on resource-constrained hardware.


\section{System Model}
\label{sec:axial_nrx}

Consider an uplink \gls{SIMO} \gls{OFDM} system. At the transmitter, an input bitstream is \gls{LDPC} encoded, mapped to symbols, and arranged into a \gls{RG} spanning $T$ \gls{OFDM} symbols and $F$ subcarriers. The resources within this grid are indexed by the symbol index $n$ and the subcarrier index $k$. \Glspl{DMRS} are transmitted at known time–frequency locations to facilitate channel estimation. After applying the \gls{IFFT}, the signal is transmitted over a 3GPP \gls{CDL} channel.

At the receiver, after synchronization and cyclic prefix removal, the \gls{FFT} is applied to each \gls{OFDM} symbol. The received signal at symbol $n$ and subcarrier $k$ is given by
\begin{equation}
  \mathbf{y}_{n,k} = \mathbf{h}_{n,k}\, x_{n,k} + \mathbf{n}_{n,k},
\end{equation}
where $\mathbf{y}_{n,k} \in \mathbb{C}^{N_\text{Rx}\times 1}$ is the received signal vector, $\mathbf{h}_{n,k} \in \mathbb{C}^{N_\text{Rx}\times 1}$ is the true channel frequency response, and $x_{n,k}$ is the transmitted symbol, normalized such that $\mathbb{E}[|x_{n,k}|^2]=1$. The term $\mathbf{n}_{n,k} \sim \mathcal{CN}(\mathbf{0},\sigma^2 \mathbf{I}_{N_\text{Rx}})$ represents the additive white Gaussian noise vector, where $N_\text{Rx}$ denotes the number of receive antennas.

\section{Neural Receiver Framework}
\label{sec:nrx_framework}
We define the neural receiver as a parameterized function
$\mathcal{F}_{\boldsymbol{\theta}}$ that maps the post-\gls{FFT} resource
grid $\mathbf{Y}$ directly to the predicted \glspl{LLR}, denoted as
$\hat{L}$ (i.e., soft-output detection):
$\hat{L} = \mathcal{F}_{\boldsymbol{\theta}}(\mathbf{Y})$.
Rather than optimizing separate and modular components for channel
estimation, equalization, and demapping, the architecture is trained
end-to-end to jointly learn this entire signal processing chain. We first
define the binary cross-entropy (BCE) loss between the ground-truth coded
bits $B \in \{0, 1\}$ and the predicted \glspl{LLR}, computed as an
empirical average over $N_{\text{tr}}$ training samples (bits/LLR pairs), as:
\begin{equation}
\label{eq:bce_loss}
\tilde{\mathcal{L}}(\boldsymbol{\theta}) =
-\frac{1}{N_{\text{tr}}}\sum_{n=1}^{N_{\text{tr}}}
\Bigl[B_n \log \sigma(\hat{L}_n)
+ (1-B_n)\log\bigl(1-\sigma(\hat{L}_n)\bigr)\Bigr],
\end{equation}
where $\sigma(\cdot)$ denotes the sigmoid activation and
$\hat{L}_n = \mathcal{F}_{\boldsymbol{\theta}}(\mathbf{Y}_n)$.
To align with communication metrics, we maximize a differentiable rate
surrogate defined in bits as
$R(\boldsymbol{\theta}) =
1-\frac{\tilde{\mathcal{L}}(\boldsymbol{\theta})}{\log(2)}$.

To prevent overfitting, the final optimization objective minimizes the
negative achievable rate surrogate alongside an $\ell_2$ weight
regularization term:
\begin{equation}
\label{eq:final_loss}
\mathcal{L}(\boldsymbol{\theta}) =
-R(\boldsymbol{\theta}) + \lambda \|\boldsymbol{\theta}\|_2^2,
\end{equation}
where $\lambda$ controls the regularization level. The details of the
axial neural receiver training are summarized in
Algorithm~\ref{alg:training}.
Furthermore, the training procedure involves periodically alternating
among different \gls{CDL} channel models to promote robust generalization
across diverse propagation conditions.

\begin{figure}
    \centering
    \includegraphics[width=0.7\linewidth]{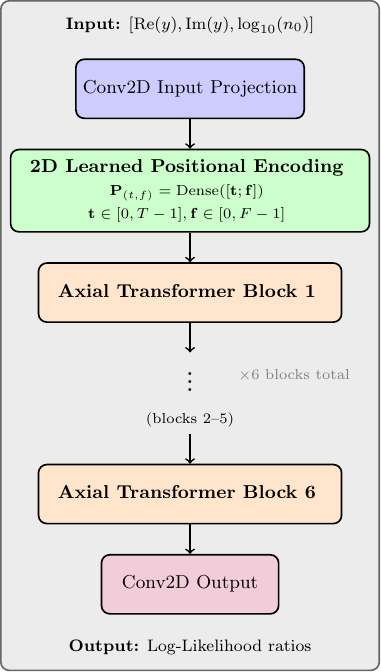}
    \caption{Architecture of proposed axial-attention transformer-based neural receiver. It comprises a 2D convolutional input projection, 2D learned positional encoding, six transformer blocks, and a 2D convolutional output projection.}
    \label{fig:axial_nrx}
\end{figure}

\begin{figure}
    \centering
    \includegraphics[width=0.7\linewidth,]{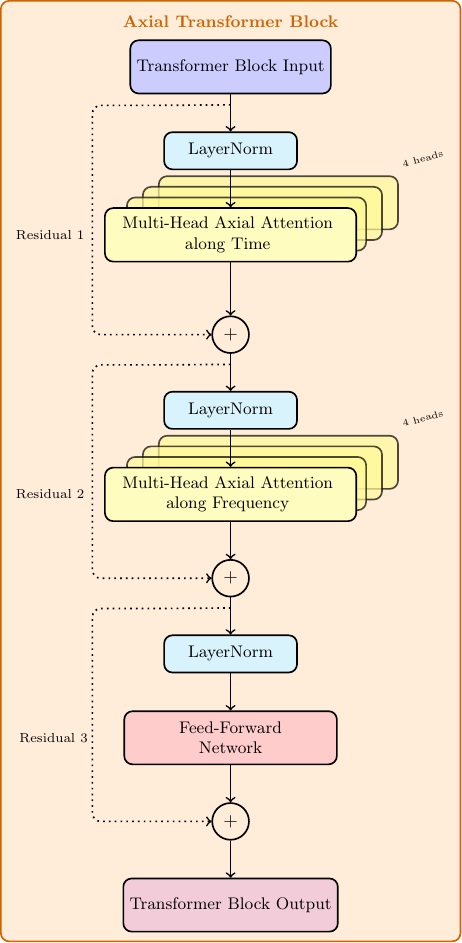}
    \caption{Axial transformer block with sequential time-axis and frequency-axis multi-head attention preceded by layer normalization. Factorized attention operations reduce computational complexity while maintaining long-range dependency modeling through residual connections.}
    \label{fig:axial_transformer}
\end{figure}

\begin{algorithm}[hb]
\small
\SetAlgoLined
\DontPrintSemicolon
\caption{Axial Neural Receiver Training Procedure}
\label{alg:training}

\KwIn{$\mathcal{C}_{\text{train}}$ (CDL Channels), $\eta$ (Learning Rate),
$N_{\text{iter}}$ (Total Training Iterations), $\lambda$ (Regularization Factor)}

\KwOut{Trained Neural Receiver $\mathcal{F}_{\boldsymbol{\theta}}$}

\textbf{Initialize:} Neural Receiver
$\mathcal{F}_{\boldsymbol{\theta}}$ with weights $\boldsymbol{\theta}$\;

$C \leftarrow \text{Sample}(\mathcal{C}_{\text{train}})$
\tcp*{Sample initial CDL model}

\For{$i = 1$ to $N_{\text{iter}}$}{
    \If{$i \mod 500 = 0$}{
        $C \leftarrow \text{Sample}(\mathcal{C}_{\text{train}})$
        \tcp*{Sample CDL model}
    }

    $E_b/N_0 \sim \mathcal{U}(E_b/N_0^{\min}, E_b/N_0^{\max})$
    \tcp*{Sample SNR}

    $\mathcal{B} \leftarrow
    \{(\mathbf{Y}_n, B_n)\}_{n=1}^{N_{\text{tr}}}
    \sim \text{System}(E_b/N_0, C)$

        $\mathbf{g} \leftarrow
    \nabla_{\boldsymbol{\theta}}
    \mathcal{L}(\boldsymbol{\theta}; \mathcal{B})$\;
    \tcp*{Compute stochastic gradient}

    $\boldsymbol{\theta} \leftarrow
    \text{Adam}(\boldsymbol{\theta},
    \text{Clip}(\mathbf{g}, 0.5), \eta)$\;
    \tcp*{Update parameters}
}

\Return $\mathcal{F}_{\boldsymbol{\theta}}$\;
\end{algorithm}

\section{Axial Attention Architecture for Neural Receiver Design}
\label{sec:axial_design}
This section focuses on the proposed axial attention transformer based neural receiver. It is designed to efficiently process a \gls{RG} of \(T\) OFDM symbols and \(F\) subcarriers to predict \glspl{LLR} \(\hat{L}\). As shown in Fig.~\ref{fig:axial_nrx}, the architecture comprises a \gls{2D} convolutional input projection, learned positional encoding, a stack of six transformer blocks, and a \gls{2D} convolutional output projection. In the following subsections, we elaborate on the specific components of the proposed architecture, and analyze their complexity.

\begin{table}[h]
\centering
\caption{Simulation Parameters for Training and Testing}
\label{tab:sim_parameters}
\footnotesize
\begin{tabular}{l p{2.2cm} p{2.2cm}}
\toprule
\textbf{Parameter} & \textbf{Training Phase} & \textbf{Testing Phase} \\
\midrule
\addlinespace[0.5em]
\multicolumn{3}{c}{\textit{Channel \& Environment}} \\
\addlinespace[0.3em]
Channel Model & CDL-\{A, B, E\} & CDL-\{C, D\} \\
Velocity & $0$--$50$ m/s (Uniform) & \parbox[t]{3.2cm}{Low: $0$--$5.1$ m/s \\ Med: $10$--$20$ m/s \\ High: $25$--$40$ m/s} \\
SNR($E_b/N_0$) & $0$--$15$ dB & $0$--$12$ dB \\
RMS Delay Spread & $10$--$100$ ns & -- \\
\midrule
\addlinespace[0.5em]
\multicolumn{3}{c}{\textit{System Configuration (Common)}} \\
\addlinespace[0.3em]
Resource Grid & \multicolumn{2}{c}{(76, 128) Subcarriers $\times$ 14 OFDM Symbols} \\
Carrier Frequency & \multicolumn{2}{c}{3.5 GHz (SCS: 30 kHz)} \\
Antenna Config. & \multicolumn{2}{c}{$N_\text{rx} = 2$} \\
Modulation & \multicolumn{2}{c}{64-QAM (Code Rate: 0.5,0.67)} \\
DMRS Config. & \multicolumn{2}{c}{Symbols 3 and 12} \\
Optimizer & \multicolumn{2}{l}{\hspace{5mm}Adam with a learning rate($\eta$):$1e^{-4}$} \\
\bottomrule
\end{tabular}
\end{table}

\subsection{Convolutional 2D Input Projection}
\label{subsec:input_projection}

The complex-valued input tensor \(\mathbf{Y} \in \mathbb{C}^{T \times F \times N_{\text{Rx}}}\) is decomposed into real ($\Re$) and imaginary ($\Im$) parts, concatenated with the noise power estimate \(N_0\):
\begin{equation}
\mathbf{Z} = \left[ \Re(\mathbf{Y}), \Im(\mathbf{Y}), \log_{10}(N_0) \cdot \mathbf{1}_{T \times F \times 1} \right] \in \mathbb{R}^{T \times F \times (2N_{\text{Rx}} + 1)}.
\label{eq:input_representation}
\end{equation}
A \gls{2D} convolutional layer projects \(\mathbf{Z}\) into embedding space \(\mathbb{R}^D\):
\begin{equation}
\mathrm{Conv2D}(\mathbf{Z}): \mathbb{R}^{T \times F \times (2N_{\text{Rx}} + 1)} \rightarrow \mathbb{R}^{T \times F \times D},
\label{eq:conv_projection}
\end{equation}
where \(D\) is the embedding dimension and the output is $\mathbf{X}_{\text{conv}} \in \mathbb{R}^{T \times F \times D}$. Unlike linear embeddings in sequence models, this 2D convolution exploits local spatial structure from channel coherence and spectral correlation, mapping each position \((t, f)\) to a \(D\)-dimensional latent vector using its local neighborhood.

\subsection{Learned Positional Encoding}
\label{subsec:positional_encoding}

Transformer architectures are inherently permutation-invariant, requiring explicit positional information to distinguish spatial locations in the time-frequency grid. We employ a trainable positional encoding \textcolor{black}{tensor $\mathbf{P} \in \mathbb{R}^{T \times F \times D}$, with one free parameter per grid location, optimized jointly with the network,} added to the convolutional projection's latent space:
\begin{equation}
\mathbf{X} = \mathbf{X}_{\text{conv}} + \mathbf{P} \in \mathbb{R}^{T \times F \times D}.
\label{eq:positional_encoding}
\end{equation}
Learned positional encodings of this form have been shown effective for transformer-based OFDM receivers~\cite{positonal_Ecoding_justify_2}. The resulting tensor $\mathbf{X}$ serves as input to the transformer blocks.

\begin{figure*}[!t]
    \centering
    \includegraphics[width=1 \linewidth]{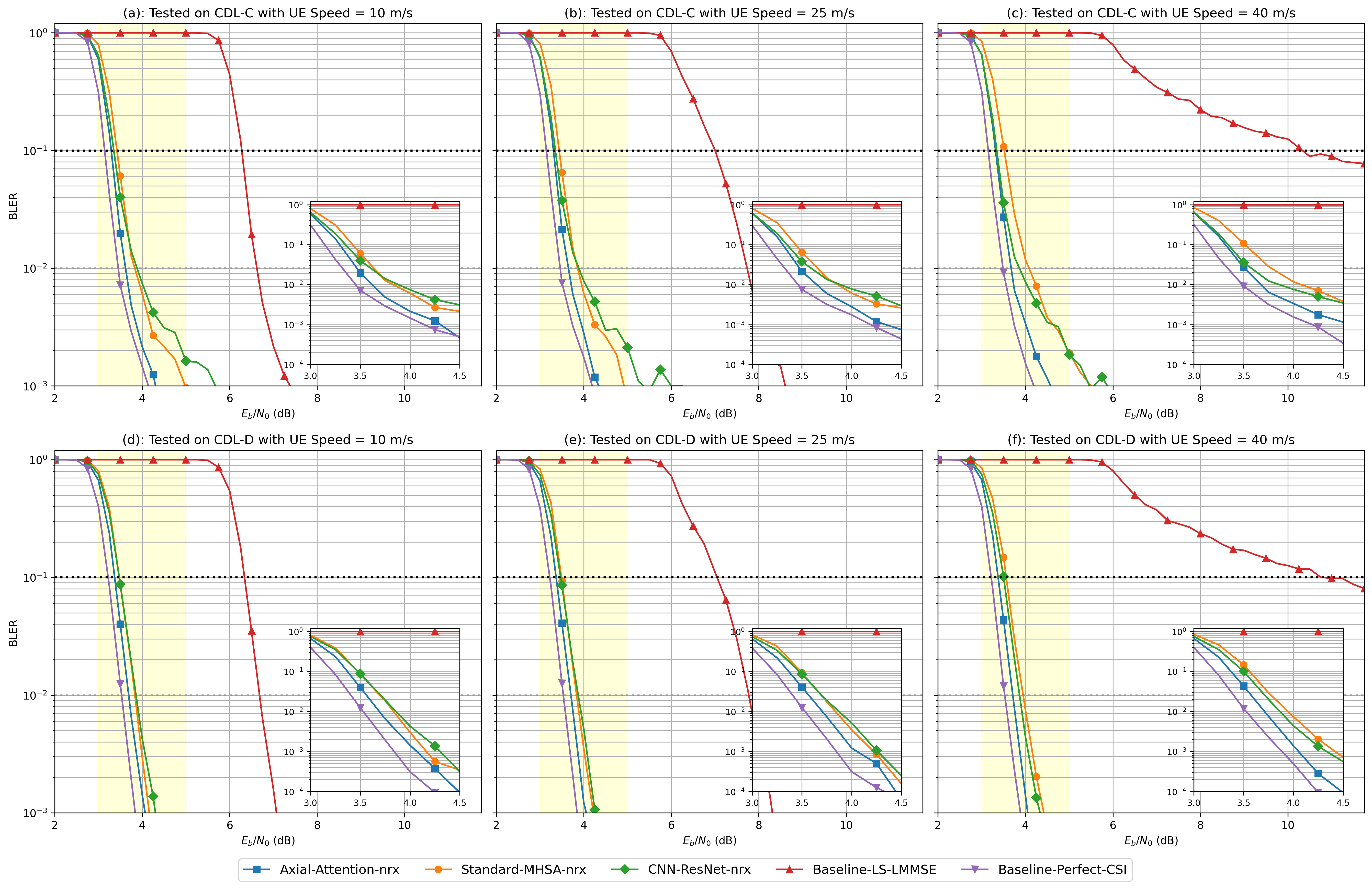}
    \caption{BLER performance under \gls{CDL}-C \gls{NLOS} and \gls{CDL}-D \gls{LOS} channels at user velocities 10--40~m/s.}
\label{fig:BLER}
\end{figure*}

\subsection{Axial Self-Attention Mechanism}
\label{subsec:axial_attention}

The attention mechanism operates on the positionally-encoded tensor \(\mathbf{X}\in\mathbb{R}^{T\times F\times D}\), decomposed into \(H\) heads with dimension \(d_h=D/H\). We first project \(\mathbf{X}\) into query, key, and value representations via learnable weights \(\mathbf{W}_{\{Q,K,V\}}^{(h)} \in\mathbb{R}^{D\times d_h}\):
\begin{equation}
    \mathbf{Q}^{(h)} = \mathbf{X} \mathbf{W}_Q^{(h)}, \quad
    \mathbf{K}^{(h)} = \mathbf{X} \mathbf{W}_K^{(h)}, \quad
    \mathbf{V}^{(h)} = \mathbf{X} \mathbf{W}_V^{(h)}. \label{eq:qkv_proj}
\end{equation}
Exploiting the separable 2D structure of OFDM grids, we factorize the global attention on these projections into sequential operations. We denote \(\mathbf{Q}^{(h)}_{\cdot,f}\) as the \(T \times d_h\) slice \textit{along the time axis} for subcarrier \(f\), and \(\mathbf{Q}^{(h)}_{t,\cdot}\) as the \(F \times d_h\) slice \textit{along the frequency axis} for symbol \(t\) (applying analogously to \(\mathbf{K}^{(h)}, \mathbf{V}^{(h)}\)).

\textit{Time-Axis Attention.} For each subcarrier \(f\in\{1,\ldots,F\}\), time-axis attention processes slices \(\mathbf{Q}^{(h)}_{\cdot,f}, \mathbf{K}^{(h)}_{\cdot,f}, \mathbf{V}^{(h)}_{\cdot,f}\in\mathbb{R}^{T\times d_h}\) as follows:
\begin{align}
    \mathbf{A}^{(h)}_{\mathrm{time},f} &=
    \mathrm{softmax}\!\left(
        \frac{\mathbf{Q}^{(h)}_{\cdot,f}(\mathbf{K}^{(h)}_{\cdot,f})^{\top}}{\sqrt{d_h}}
    \right)\in\mathbb{R}^{T\times T}, \label{eq:time_A}\\
    \mathbf{Y}^{(h)}_{\mathrm{time},f} &=
    \mathbf{A}^{(h)}_{\mathrm{time},f}\mathbf{V}^{(h)}_{\cdot,f}
    \in\mathbb{R}^{T\times d_h}. \label{eq:time_Yh}
\end{align}
Multi-head aggregation via concatenation and linear projection with learnable output matrix \(\mathbf{W}_O\in\mathbb{R}^{D\times D}\) yields

{\small
\begin{equation}
\mathrm{Att}_{\mathrm{time}}(\mathbf{X})_{\cdot,f} = \mathrm{Concat}\big(\mathbf{Y}^{(1)}_{\mathrm{time},f},\ldots,\mathbf{Y}^{(H)}_{\mathrm{time},f}\big)\mathbf{W}_O \in\mathbb{R}^{T\times D}.
\label{eq:time_agg}
\end{equation}
}
Stacking across all \(F\) subcarriers produces \(\mathrm{Att}_{\mathrm{time}}(\mathbf{X})\in\mathbb{R}^{T\times F\times D}\).

\textit{Frequency-Axis Attention.} Analogously, for each OFDM symbol \(t\in\{1,\ldots,T\}\), frequency-axis attention operates on slices \(\mathbf{Q}^{(h)}_{t,\cdot}, \mathbf{K}^{(h)}_{t,\cdot}, \mathbf{V}^{(h)}_{t,\cdot}\in\mathbb{R}^{F\times d_h}\) as follows:
\begin{align}
    \mathbf{A}^{(h)}_{\mathrm{freq},t} &=
    \mathrm{softmax}\!\left(
        \frac{\mathbf{Q}^{(h)}_{t,\cdot}(\mathbf{K}^{(h)}_{t,\cdot})^{\top}}{\sqrt{d_h}}
    \right)\in\mathbb{R}^{F\times F}, \label{eq:freq_A}\\
    \mathbf{Y}^{(h)}_{\mathrm{freq},t} &=
    \mathbf{A}^{(h)}_{\mathrm{freq},t}\mathbf{V}^{(h)}_{t,\cdot}
    \in\mathbb{R}^{F\times d_h}. \label{eq:freq_Yh}
\end{align}
Multi-head aggregation yields
{\small
\begin{equation}
\mathrm{Att}_{\mathrm{freq}}(\mathbf{X})_{t,\cdot} = \mathrm{Concat}\big(\mathbf{Y}^{(1)}_{\mathrm{freq},t},\ldots,\mathbf{Y}^{(H)}_{\mathrm{freq},t}\big)\mathbf{W}_O \in\mathbb{R}^{F\times D},
\label{eq:freq_agg}
\end{equation}
}
with stacking producing \(\mathrm{Att}_{\mathrm{freq}}(\mathbf{X})\in\mathbb{R}^{T\times F\times D}\).

\textit{Sequential Composition.} The axial transformer block applies both operations sequentially with residual connections:
\begin{align}
    \mathbf{X} &\leftarrow \mathbf{X} + \mathrm{Att}_{\mathrm{time}}(\mathbf{X}), \label{eq:res_time}\\
    \mathbf{X} &\leftarrow \mathbf{X} + \mathrm{Att}_{\mathrm{freq}}(\mathbf{X}). \label{eq:res_freq}
\end{align}

Time-axis attention captures temporal dependencies among OFDM symbols, subsequently refined by frequency-axis attention modeling spectral correlations.
\begin{figure}[H]
    \centerline{
    \includegraphics[width=1\linewidth]{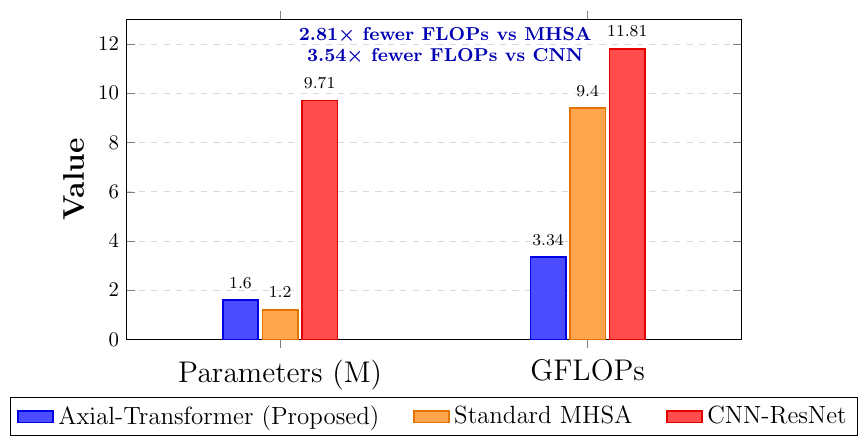}}
    \caption{Model complexity of neural receiver architectures}
    \label{fig:model_complexity}
\end{figure}

\section{Model Efficiency and Numerical Results}
\label{sec:efficiency_and_results}
To assess the computational efficiency of the proposed axial attention architecture, we benchmark it against both a global \gls{MHSA}-based receiver and a \gls{CNN}-ResNet baseline. The axial and global attention approaches share the same end-to-end structure described in Section~\ref{sec:axial_design}; \emph{the difference lies solely in the attention mechanism within the transformer blocks.} 

As discussed in Section~\ref{sec:intro}, standard global MHSA exhibits a quadratic complexity of $\mathcal{O}(T^2F^2)$. By factorizing the pairwise similarity computation along the temporal and spectral dimensions, the proposed axial attention limits this complexity to $\mathcal{O}(TF(T+F))$. For the representative 5G NR parameters evaluated here ($T=14$, $F=128$), this theoretical advantage translates to a complexity reduction factor of $\frac{TF}{T+F} \approx 12.6\times$ relative to global attention.

\textbf{Model Efficiency:} As illustrated in Fig.~\ref{fig:model_complexity}, the axial architecture achieves substantial computational savings over both baselines while maintaining competitive parameter count. Although the factorized attention mechanism requires separate projection matrices for time and frequency axes, increasing parameters by $1.3\times$ relative to standard MHSA, this modest overhead enables a $2.81\times$ reduction in FLOPs. The resulting efficiency makes the axial receiver suitable for resource-constrained 6G edge deployments.

All receiver architectures (axial attention, global \gls{MHSA}, and \gls{CNN}-ResNet) are trained end-to-end to map received resource grids to \glspl{LLR}, using an identical regularization scheme and optimization hyperparameters to ensure a fair comparison, the full configuration is summarized in Table~\ref{tab:sim_parameters}\footnote{CNN-ResNet needed further fine-tuning}. We further benchmark the proposed axial receiver against LS channel estimation-LMMSE equalization and an ideal receiver with perfect CSI. Training is performed on NVIDIA A40 GPUs, and all simulations are implemented in Sionna~\cite{sionna}. Figure~\ref{fig:BLER} reports the resulting BLER across UE velocities under NLOS (CDL-C) and LOS (CDL-D) channel conditions 

\textbf{Performance Analysis:} Under NLOS conditions (Fig.~\ref{fig:BLER}, top), the axial receiver consistently outperforms all baselines. At $1\%$ BLER, it achieves SNR gains of $0.25$--$0.40~\mathrm{dB}$ over standard MHSA and $0.20$--$0.30~\mathrm{dB}$ over CNN-ResNet. Notably, LS-LMMSE fails to reach $1\%$ BLER at $40~\mathrm{m/s}$ due to rapid channel variation, while the axial receiver maintains robust performance at $3.70~\mathrm{dB}$ SNR. Similar trends hold for LOS conditions (Fig.~\ref{fig:BLER}, bottom), where the axial architecture maintains a $0.15$--$0.25~\mathrm{dB}$ SNR gain over neural baselines at $1\%$ BLER and outperforms LS-LMMSE by margins exceeding $7~\mathrm{dB}$ at high mobility. These results confirm the architecture's superior ability to capture temporal dependencies essential for high-mobility tracking.

\section{Conclusion and Future Work}
\label{sec:conclusion}
This work proposes axial attention as a computationally efficient framework for neural receivers in AI-native 6G systems. By factorizing self-attention along temporal and spectral dimensions, the architecture overcomes the quadratic scalability bottleneck of conventional transformers while retaining the global context of the \gls{RG}. Our results demonstrate that the axial receiver achieves consistent performance gains over CNN and LS baselines, particularly at stringent 1\% BLER targets while reducing inference GFLOPs by over $3.5\times$ compared to CNNs. These properties make axial attention based architectures excellent candidate for resource-constrained edge deployments requiring ultra-reliable low-latency processing.
Future work on axial attention architecture will focus on two key directions, namely extension to MIMO configurations and low-bit quantization.

\nocite{kingma2014adam,a-mmse,nir_bayesian_neural_mimo}
\bibliographystyle{IEEEbib}
\bibliography{references}

@book{molisch2011wireless,
    author = "A. F. Molisch",
    title = "Wireless Communications",
    publisher = "John Wiley \& Sons",
    year = 2011,
    note = "Chapter 6: Wideband and Directional Channel Characterization",
    url = "https://www.wiley.com/en-us/Wireless+Communications%2C+2nd+Edition-p-9780470741870"
}

@INPROCEEDINGS{positonal_Ecoding_justify_2,
  author={Guler, B. and Jafarkhani, H.},
  booktitle={ICC 2025 - IEEE International Conference on Communications}, 
  title={{AdaFortiTran: An Adaptive Transformer Model for Robust OFDM Channel Estimation}}, 
  year={2025},
  volume={},
  number={},
  pages={3797-3802},
  doi={10.1109/ICC52391.2025.11160810}}

@INPROCEEDINGS{saketh_asilomar25,
  title={{Efficient Deep Neural Receiver with Post-Training Quantization}},
  booktitle={IEEE 59th Asilomar Conference on Signals, Systems, and Computers},
  author={Yellapragada, S. S. and Ollila, E. and Costa, M.},
  journal={arXiv preprint arXiv:2508.06275},
  year={2025}
}

@INPROCEEDINGS{saketh_qat,
  title={{Efficient Quantization-Aware Neural Receivers: Beyond Post-Training Quantization}},
  booktitle={IEEE International Conference on Acoustics, Speech and Signal Processing (ICASSP)},
  author={Yellapragada, S. S. and Ollila, E. and Costa, M.},
  journal={arXiv preprint arXiv:2509.13786},
  year={2026}
}

@article{unified_transformer,
  title={{A Unified Transformer Architecture for Low-Latency and Scalable Wireless Signal Processing}},
  author={Kawai, Y. and Koodli, R.},
  journal={https://arxiv.org/pdf/2508.17960},
  year={2025}
}

@misc{a-mmse,
      title={{Attention-Aided MMSE for OFDM Channel Estimation: Learning Linear Filters with Attention}}, 
      author={T. Ha and C. Jung and H. Kim and J. Park and J. Park},
      year={2026},
      eprint={2506.00452},
      archivePrefix={arXiv},
      primaryClass={eess.SP},
      url={https://arxiv.org/abs/2506.00452},
      howpublished = {arXiv preprint}
}

@ARTICLE{deeprx,
  author={Honkala, M. and Korpi, D. and Huttunen, J.},
  journal={IEEE Transactions on Wireless Communications}, 
  title={{DeepRx: Fully Convolutional Deep Learning Receiver}}, 
  year={2021},
  volume={20},
  number={6},
  pages={3925-3940},
  doi={10.1109/TWC.2021.3054520}}

@INPROCEEDINGS{deeprx_mimo_icc,
  author={Korpi, D. and Honkala, M. and Huttunen, J. and Starck, V.},
  booktitle={IEEE International Conference on Communications}, 
  title={{DeepRx MIMO: Convolutional MIMO Detection with Learned Multiplicative Transformations}}, 
  year={2021},
  volume={},
  number={},
  pages={1-7},
  doi={10.1109/ICC42927.2021.9500518}}

@INPROCEEDINGS{nrx_seb,
  author={Cammerer, S. and Aoudia, F. A. and Hoydis, J. and Oeldemann, A. and Roessler, A. and Mayer, T. and Keller, A.},
  booktitle={IEEE Globecom Workshops}, 
  title={{A Neural Receiver for 5G NR Multi-User MIMO}}, 
  year={2023},
  volume={},
  number={},
  doi={10.1109/GCWkshps58843.2023.10464486}}

@ARTICLE{trainable_seb,
  author={Cammerer, S. and Aoudia, F. A. and Dörner, S. and Stark, M. and Hoydis, J. and ten Brink, S.},
  journal={IEEE Trans. Commun.}, 
  title={Trainable Communication Systems: Concepts and Prototype}, 
  year={2020},
  volume={68},
  number={9},
  pages={5489-5503},
  doi={10.1109/TCOMM.2020.3002915}}

@article{ho2019axial,
  title        = {Axial Attention in Multidimensional Transformers},
  author       = {Ho, J. and Kalchbrenner, N. and Weissenborn, D. and Salimans, T.},
  year         = {2019},
  eprint       = {1912.12180},
  archivePrefix= {arXiv},
  primaryClass = {cs.CV},
  doi          = {10.48550/arXiv.1912.12180},
  journal          = {https://arxiv.org/abs/1912.12180}
}

@inproceedings{axialpanoptic,
author = {Wang, H. and Zhu, Y. and Green, B. and Adam, H. and Yuille, A. and Chen, L.},
title = {{Axial-DeepLab: Stand-Alone Axial-Attention for Panoptic Segmentation}},
year = {2020},
booktitle = {Proceedings of the 16th European Conference on Computer Vision},
pages = {108–126},
location = {}
}

@ARTICLE{atchutJournal,
  author={Kocharlakota, A. K. and Vorobyov, S. A. and Heath, R. W.},
  journal={IEEE Transactions on Wireless Communications}, 
  title={Pilot Contamination Aware Transformer for Downlink Power Control in Cell-Free Massive MIMO Networks}, 
  year={2026},
  volume={25},
  number={},
  pages={9656-9671},
  doi={10.1109/TWC.2025.3643786}}

@ARTICLE{tingtingGAN,
  author={T Zhang and S. A. Vorobyov and D. J. Love and T. Kim and K. Dong},
  journal={IEEE Wireless Communications Letters}, 
  title={Pilot Contamination-Aware Graph Attention Network for Power Control in CFmMIMO}, 
  year={2026},
  volume={15},
  number={},
  pages={1464-1468},
  doi={10.1109/LWC.2026.3656252}}

@software{sionna,
 title = {Sionna},
author = {Hoydis, J. and Cammerer, S. and Aoudia, F. A. and
 Nimier-David, M. and Maggi, L. and Marcus, G. and Vem, A. and Keller,
 A.},
 note = {https://nvlabs.github.io/sionna/},
 year = {2022},
 version = {1.1.0}
}

@ARTICLE{e2e_faa,
  author={Ait Aoudia, F. and Hoydis, J.},
  journal={IEEE Transactions on Wireless Communications}, 
  title={{End-to-End Learning for OFDM: From Neural Receivers to Pilotless Communication}}, 
  year={2022},
  volume={21},
  number={2},
  pages={1049-1063},
  doi={10.1109/TWC.2021.3101364}}

@ARTICLE{nir_bayesian_neural_mimo,
  author={Raviv, T. and Park, S. and Simeone, O. and Shlezinger, N.},
  journal={IEEE Transactions on Vehicular Technology}, 
  title={{Uncertainty-Aware and Reliable Neural MIMO Receivers via Modular Bayesian Deep Learning}}, 
  year={2025},
  volume={74},
  number={11},
  pages={17637-17651},
  doi={10.1109/TVT.2025.3579989}}

@inproceedings{vaswani2017attention,
  title={{Attention is all you need}},
  author={Vaswani, A. and Shazeer, N. and Parmar, N. and Uszkoreit, J. and Jones, L. and Gomez, A. N. and Kaiser, {\L}. and Polosukhin, I.},
  booktitle={Proceedings of the 31st International Conference on Neural Information Processing Systems},
  pages={6000--6010},
  year={2017}
}

@inproceedings{kingma2014adam,
	author        = {Kingma, D. P. and Ba, J.},
	booktitle     = {Proceedings of the International Conference on Learning Representations},
	numpages      = {13},
	title         = {{Adam: A Method for Stochastic Optimization}},
	year          = {2015},
}

\end{document}